\documentclass[preprint,showpacs,preprintnumbers,amsmath,amssymb,floats]{revtex4}
\usepackage[dvips]{graphicx}
\usepackage[english]{babel}
\usepackage{amsmath}
\usepackage{amssymb}
\usepackage{color}
\usepackage{epstopdf}

\newcommand{\be}{\begin{eqnarray}}
\newcommand{\ee}{\end{eqnarray}}

\newcommand{\au}{a_{i,\uparrow}}
\newcommand{\ad}{a_{i,\downarrow}}

\begin{document}

\title{Quantum Criticality in Quasi-Two Dimensional Itinerant Antiferromagnets}

\author{C. M. Varma}
\affiliation{Department of Physics and Astronomy, University of California, Riverside CA 92521, USA}

\date\today
\begin{abstract}
Quasi-two dimensional itinerant fermions in the Anti-Ferro-Magnetic (AFM) quantum-critical region of their phase diagram, such as in the Fe-based superconductors or in some of the heavy-fermion compounds, exhibit a resistivity varying linearly with temperature and a contribution to specific heat or thermopower proportional to $T \ln T$. It is shown here that a generic model of itinerant AFM can be canonically transformed so that its critical fluctuations around the AFM-vector $Q$ can be obtained from the fluctuations in the long wave-length limit of a dissipative quantum XY model. The fluctuations of the dissipative quantum XY model in 2D have been evaluated recently and in a large regime of parameters, they are determined, not by renormalized spin-fluctuations but by topological excitations. In this regime, the fluctuations  are separable in their spatial and temporal dependence and have a spatial correlation length which is proportional to logarithm of the temporal correlation length, i.e. for some purposes the effective dynamic exponent $z =\infty.$  The time dependence gives $\omega/T$-scaling at criticality. The observed resistivity and entropy then follow.  Several predictions to test the theory are also given. 

\end{abstract}
\maketitle

The problem of AFM quantum-critical fluctuations  in itinerant fermions has been studied extensively \cite{Hilbert, Moriya, BealMonod, Hertz, Millis} by simple extensions of the theory of classical critical fluctuations. This idea has been proven by S-S. Lee \cite{Lee} to be uncontrolled in two dimensions. (The theory is controlled for AFM fluctuations in 3D; the measured fluctuation spectra and the 
 properties calculated \cite{nishiyama} from it agree well with the experiments also.) Lee has also proposed methods for expansion about 3 dimensions for a problem with a 1 dimensional fermi-surface, as well as a different expansion about a line in the spatial dimension - Fermi surface dimension plane. Other procedures \cite{Metlitski, Efetov, Raghu} have also been proposed, each yielding different results. While these methods (at least to linear order in the expansion parameter) appear controlled, they do not give the observed singular-Fermi-liquid properties. All these are theories of criticality due to renormalized spin-waves.  Other semi-phenomenological ideas \cite{Si, coleman, wolf}, with varying degrees of justification have also been proposed. Imaginative ideas based on string theory-duality have also been advanced \cite{zaanen}.  At least so far, there is no sense of a symmetry breaking in such theories, which appears invariably in experiments astride the region of singular Fermi-liquid properties. 

The linear in $T$ resistivity and the $T~log~T$ specific heat and thermopower in the AFM quantum-critical region in 2D \cite{Steglich} \cite{Zhou, Analytis} are reminiscent of the properties in the similar region in hole-doped cuprate superconductors.  The quantum critical point associated with the singular Fermi-liquid properties in the hole-doped cuprates is obviously not of the AFM order, which goes to 0 at dopings far from the regime of such anomalous metallic properties \cite{footnote1}.  A quite different order parameter, which does not break translational symmetry, was predicted \cite{cmv97} for which there is experimental evidence in many different kinds of experiments \cite{bourges-rev, kaminski, armitage, shekhter}. The fluctuations of such an order parameter can be mapped to a dissipative quantum XY model with four-fold anisotropy \cite{aji}. 

The observation of similar singular Fermi-liquid properties in the AFM quantum-critical region suggests an investigation to see if AFM fluctuations are also described by a similar model. 
A generic model of itinerant fermions, which have a commensurate or an incommensurate planar AFM transition, or one which has an incommensurate uni-axial transition, is shown here to transform canonically to a model with a superconductive transition, which is described by a dissipative quantum XY model. The fluctuations of the AFM model near the AFM wave-vector ${\bf Q}$ can be obtained from the known fluctuations of the XY model in the long wave-length limit. Fermions acquire the observed singular properties through scattering such fluctuations. It is generally agreed that a pre-requisite for understanding superconductivity is understanding the normal state anomalies above $T_c$. 


{\bf Canonical Transformation}: Consider the following Hamiltonian for fermions
\be
\label{mod1}
H =  \sum_{<i j>, \sigma = \uparrow, \downarrow} t_{ij} a_{i, \sigma}^{\dagger} a_{j,\sigma}  + H.C. + U \sum_i (n_{i \uparrow}-1/2) (n_{i \downarrow}-1/2)  +  I_z (S_i^z)^2  - \mu n_i + h S_i^z.
\ee
$<ij>$ sums over nearest neighbors on a bi-partite two dimensional lattice. $U >0$ so that for large enough $U/t$, a Mott insulating state is expected with AFM correlations or commensurate order at half-filling when the chemical potential $\mu = 0$. Beyond some deviation  from half-filling, a metallic state is expected, with AFM correlations at low enough temperatures. These correlations are in general peaked at the incommensurate vectors ${\bf Q} = ({\bf Q}_0 + {\bf q}_0)$ with ${\bf Q}_0.{\bf R}_0 = \pi$, where ${\bf R}_0$'s are the nearest neighbor vectors and ${\bf q}_0$ depends on the deviation from half-filling. A single ion anisotropy term with coefficient  $I_z > 0$ ensures that the AFM correlations are stronger for planar spin-correlations, i.e. spin  in the xy plane, and  $I_z < 0$ ensures the same for uni-axial correlations, i.e spins along the z-axis. Only $h=0$ is considered in this paper but finite $h$ may be useful in further work. No magnetic order is expected for large enough deviation from half-filling.  So, there is a quantum critical point as a function of doping. The Hamiltonian of Eq. (\ref{mod1}) may be paradigmatic of a general class of models with AFM correlations, but specific details of the Hamiltonian for the actual experimental systems need to be examined to be certain.

The (canonical) transformations \cite{micnas}, 
  \be
  \label{cantrans}
  \au &\to& e^{i\zeta_i}{\tilde{a}}_{i, \uparrow};  ~{{a}}_{i, \uparrow}^{\dagger} \to e^{-i\zeta_i}{\tilde{a}}_{i, \uparrow}^{\dagger}; \\ \nonumber
   \ad &\to& {\tilde{a}}_{i, \downarrow}^{\dagger} e^{iQ_0.R_i+i\zeta_i}  ; ~{a}_{i, \downarrow}^{\dagger} \to {\tilde{a}}_{i, \downarrow}e^{-iQ_0.R_i-i\zeta_i}.
  \ee
  with 
  \be
  \label{zeta}
  \zeta_i= -\frac{1}{2} {\bf q}_0\cdot {\bf R}_i,
  \ee
  transform the Hamiltonian of (\ref{mod1}) to 
  \be
  \label{mod2}
  \tilde{H} &=& - \tilde{U} \sum (\tilde{n}_{i \uparrow}-1/2)(\tilde{n}_{i \downarrow} -1/2) -  \sum_i (\tilde{h} \tilde{S}^z_{i} + \tilde{\mu} n_i )\\ \nonumber
  &+& \sum_{<i j>,~(\alpha = \pm } \tilde{{t}}_{ij} e^{-i\alpha(\zeta_i-\zeta_j)}  \tilde{a}_{i, \sigma}^{\dagger} \tilde{a}_{j,\sigma} + H.C.
  \ee
Here $\alpha = \pm$ for $\sigma = \uparrow,\downarrow$, respectively, and
  \be
  \label{parameters}
    \tilde{t} &=& t; ~~\tilde{U} = U-2I_z, \tilde{h} = \mu, \tilde{\mu} =h.
  \ee
 The transformed Hamiltonian is a model with on-site attractive interactions, a Zeeman field related to the deviation of the original model from half-filling and a spin-dependent phase factor ($\alpha(\zeta_i-\zeta_j)$, $\alpha = (\pm 1)$ for $\sigma = (\uparrow,\downarrow)$), on the link $(i,j)$ related to the incommensurate vector ${\bf q}_0$ or the deviation from half-filling. As a result, the Fermi-surface of up and down spins are shifted in 
 opposite directions by $\pm {\bf q}_0/2$; thus  $\alpha(\zeta_i-\zeta_j)$ is a spin-orbit field. Corresponding to the transitions to planar AFM and uni-axial AFM in model (\ref{mod1}), model (\ref{mod2}) has a superconducting ground state for small enough $\tilde{h}$ for $I_z > 0$ and a charge density wave for $I_z < 0$. Also, corresponding to a quantum critical point in model (\ref{mod1}) for ${\mu} = \mu_c$ with other parameters fixed, there is a quantum critical point in model (\ref{mod2}) for $\tilde{h} = \tilde{h}_c$, as will be clearer below.


 
{\bf Relation of Spin-Correlations to Superconducting Correlations}: 
With the canonical transformations, the spin-raising/lowering operator in $H$ are related to the Cooper pair creation/annihilation operator in $\tilde{H}$, and $S_i^z$ is related to the density operator, 
\be
\label{S-trans}
  S_i^+ &\to & e^{i{\bf Q}.{\bf R}_i} \Psi^+_i, ~~ S_i^- \to  e^{-i{\bf Q}.{\bf R}_i}\Psi_i ; ~~ S_i^z \to \tilde{n_i} -1 \\
  \Psi^+_i &= & \tilde{a}_{i\uparrow}^+\tilde{a}_{i\downarrow}^+, ~etc.
\ee
Define the response functions for two operators $A$ and $B$ for a Hamiltonian $H$ by 
\be
\label{resp}
\chi^H_{(AB)}(i,j;t-t') = - i \theta(t-t') \langle [A_i(t), B_j(t')]\rangle_H
\ee
Consider $I_z<0$ so that $\chi^H_{(S^zS^z)}({\bf Q + q}, \omega)$ are important. They map to incommensurate charge density fluctuations at the same momenta. Such fluctuations are described by the fluctuations of an XY model \cite{mcmillan}. This follows from the fact that an {\it incommensurate} wave of charge (or z-component of magnetization) has in general an order parameter $A \sin({\bf Q\cdot R}_i + \phi)$, where $A$ is the amplitude. Any spatially uniform value of $\phi$ has the same energy, just as the phase-variable in a superfluid. Spatial variations in $\phi$ cost an energy $\propto \rho_{s \|} |\nabla_{\|} \phi|^2 +  \rho_{s \bot}|\nabla_{\bot} \phi|^2$, where $\nabla_{\|, \bot}$ refer to variations parallel and perpendicular to ${\bf Q}$. Also the energy can only depend periodically on the difference of phase $(\phi_i-\phi_j)$  between two points $i$ and $j$ on the lattice. Therefore, the uniaxial incommensurate AFM fluctuations are described by an XY model. The edge dislocations in the incommensurate wave in 2D correspond to vortices in 2D superfluids. For the uni-axial case, unlike the case for the planar case discussed below, the mapping of Eq. (\ref{cantrans}) is in fact unnecessary. 

Consider $I_z >0$ so that the important fluctuations are planar. These are the relevant fluctuations for the Fe-based compounds and for some heavy Fermions. It follows, using the definition (\ref{resp}) that knowledge of any response function of model (\ref{mod1}) gives also a response function of (\ref{mod2}) and vice-versa. The two are related by the (\ref{cantrans}). In particular, the planar spin-response function 
in the model of Eq. (\ref{mod1}) is identical to the Cooper pair response function for the model of Eq. (\ref{mod2}):
\be
\label{trans-q}
\chi^H_{(S^+S^-)}({\bf Q + q}, \omega) \equiv \chi^{\tilde{H}}_{(\Psi^+\Psi)}({\bf q}, \omega).
\ee
The identity (\ref{trans-q}) asserts that {\bf if} the correlation function at the left diverges at $q=0$ for some parameters, signifying an AFM transition, the correlation function at the right also diverges at $q=0$ for parameters related to each other by (\ref{parameters}), signifying a uniform s-wave superconducting transition. Moreover, the planar AFM correlation at small ${\bf q}$ around ${\bf Q}$ at any $\omega$ in model (\ref{mod1}) may be obtained exactly from the superconducting correlations at ${\bf q}$ at the same $\omega$  in model (\ref{mod2}). Either model may have other phase transitions, which would also bear correspondence. They are not relevant to the problem addressed here, which has only to do with finding the correlation functions for the paramagnetic to AFM transition in model (\ref{mod1}).

The relation between the correlation functions does not say anything at all about the value of the parameters where the critical point occurs. It is however worthwhile to discuss the physical reason for the transition in the superconducting model with a Zeeman field.
The Zeeman field in model (\ref{mod2}) make the Fermi-sphere for one spin bigger than the other and the spin-orbit field displaces them with respect to each other by $2 {\bf q}_0$. The spin-orbit as well as the Zeeman field are taken into account in the one-particle spectra by the condition of equal chemical potential, by introducing spin-dependent Fermi-vectors
 \be
 \label{pf}
{\bf p}_F  &=&{\bf p}^0_F  + (\delta {\bf p}_F) \sigma_3 ; ~~ \delta {\bf p}_F \equiv {\bf q}_0  + \frac{g\mu_B\tilde{h}}{|v_F|}
\ee
for $q_0/p^0_F \ll 1$. 
Time-reversal symmetry is preserved by the shift ${\bf q}_0\sigma_3$ while it is broken by the shift proportional to $\tilde{h}$. The latter leads to a displacement  in momentum of the up and down Fermi-surfaces. Therefore the usual logarithmic singularity for s-wave Cooper pairing at  zero total momentum (q=0), due to attractive interactions, 
is cut-off due to the spin-splitting energy $g\mu_B \tilde{h}$. There is no transition even at $T\to 0$ for ${\tilde h}$ larger than a critical field 
${\tilde h}_c$. This corresponds to the AFM quantum-critical point in repulsive U model at a critical value $\mu_c$ connected to ${\tilde h}_c$ by (\ref{parameters}). 

The approach to finding the quantum-critical correlations of the itinerant AFM in 2d, by using Eq. (\ref{trans-q}), is worthwhile because the quantum-critical correlations of the superconductor in 2d are known rather accurately \cite{zhuchencmv}.
Near the phase transitions of model (\ref{mod2}), we may, using techniques such as the Hubbard-Stratonovich transformation, write it in terms of a Hamiltonian for its collective fluctuations $H_{coll}$, for the Fermions $H_F$ and for the interaction between the fermions and the collective fluctuations $H_{int}$.
\be
H = H_F + H_{coll} + H_{int}.
\ee
The model for collective critical fluctuations in a superconductor may be expressed in terms of the pair-field operators $\Psi$, which are products of a pair of time-reversed fermions. In 2D, the amplitude fluctuations are irrelevant and the phase fluctuations determine the critical properties. The critical fluctuations are then those for an XY model for a field $\Psi({\bf r}, \tau) \equiv |\Psi| e^{i \theta({\bf r}, \tau)}$, with $|\Psi|$ weakly enough varying that it may be kept fixed \cite{KT, chakra-kivel}. The action for $H_{coll}$ for the 2d-XY model, with a four-fold anisotropy term and including dissipation, is expressed in terms of the phase $\theta_i(\tau)$ on a lattice of sites ${\bf R}_i$ as, 
\be
\label{mod-xy}
S_{coll} = -\int_0^{\beta}d\tau \sum_i \frac{1}{2E_c} \Big(\frac{d\theta_i(\tau)}{d\tau}\Big)^2 + K_0 \sum_{j(i)} \cos\big(\theta_i(\tau) - \theta_j(\tau)\big) + h_4 \cos 4\theta_i(\tau) + S_{diss}.
\ee
The relationship of the parameters in (\ref{mod-xy}) and (\ref{mod2}) is hard to derive microscopically, except for weak-coupling or for strong coupling, $|U|/t << 1$, or $ >> 1$, respectively. In general terms, $K$ is related to the superfluid density which decreases as the Zeeman field $\tilde{h}$ increases, and $E_c$ to the compressibility. $h_4$ reflects the anisotropy of the kinetic energy parameter $t_{ij}$. The relations locate the quantum-critical point but they are unnecessary for finding the correlation functions around the critical point. 

$S_{diss}$ is the dissipative term in the action. It is necessary to show that, under the transformations (\ref{S-trans}), the form of the dissipation also goes from that in one model to that of the other. The dissipation used \cite{Moriya, Hertz} in the itinerant AFM on symmetry grounds is of the form
\be
\label{af-diss}
S_{diss} = \sum_{\omega,{\bf q}}  i \alpha |\omega| |{S}({\bf Q+q}, \omega)|^2. 
\ee
This arises from decay of collective AFM spin-fluctuations into incoherent particle-hole pairs with spin 1.  In the problem of quantum-criticality of the XY model \cite{chakra-kivel}, the nature of dissipation has been chosen to be that of the Caldeira-Leggett form \cite{Caldeira1983}, which is due to the decay of collective super-current ${\bf J}$ to incoherent fermion current. 
The current ${\bf J}$ is proportional to the gradient of the phase, ${\bf \nabla}~ \theta$, so that the Caldera-Leggett dissipation for small ${\bf q}$ is, 
\be
\label{sc-diss}
S^{CL}_{diss} = \sum_{{\bf q}, \omega} i~\alpha' |\omega| ~q^2 ~|\theta(q,\omega)|^2.
\ee 
Here ${\alpha'} = \frac{1}{4\pi^2} R_Q/R_s$; $R_Q$ is the quantum of resistance for Cooper pairs, equal to $h/4e^2$ and $R_s$ is the resistance per square of the normal state \cite{chakra-kivel}.  Under the transformations (\ref{sc-diss}) the super-current operator ${\bf J}_{ij} \propto  Im (\Psi_i^+ \Psi_j)$  transforms to  $Im (S^+_i S^-_j e^{i{\bf Q \cdot R}_{ij}})$. On Fourier transformation, this becomes $|{\bf Q + q}|^2 Im{S^+S^-}({\bf Q+q}, \omega)$. $q$ may be dropped in $|{\bf Q + q}|^2$ because of the large fixed $|Q|$. In 2D, only the imaginary part of the order parameter correlations are critical. It follows that the Caldera-Leggett dissipation (\ref{sc-diss}), leads on using the transformations (\ref{S-trans}), to the usual dissipation of the itinerant AFM model (\ref{af-diss}) with $\alpha = \alpha'|{\bf Q}|^2$. Similar proportionality for dissipation for the phase fluctuations of the incommensurate uni-axial model to dissipation in the XY model also follows.

The dissipative quantum 2D-XY model has a rich phase diagram \cite{zhuchencmv, zhuchencmv2, Sudbo} at $T=0$. At $\alpha=0$, it has a transition of the 3D-XY class for $E_c/K_0 \lesssim 12$ with the dynamical critical exponent $z=1$. As $\alpha$ increases, the transition continues to be in same class with the critical ratio of $E_c/K_0$ increasing slightly, till about $\alpha \approx 0.01$, beyond which, it changes to the 
$z=\infty$, with the critical value of $E_c/K_0$ sharply increasing with the critical value of $\alpha$. The model also has some interesting cross-overs to 2D critical behavior of the Kosterlitz-Thouless kind and from that to the 3D ordered state as a function of $T^2/(K_0E_c)$. We focus here on the $T=0$ quantum critical response at the disordered to the 3D ordered phase transition with dynamical critical exponent $z \to \infty$, as it appears to be relevant to the experiments. It is important to note that this occupies a substantial part of the phase diagram. This may be seen from the fact that $\alpha$ is proportional to the inverse 2D resistivity and its lower limit is bounded by the maximum resistivity possible  for a disordered 2D problem to be considered itinerant.  $z=1$ transition only occurs for the very disordered problem with resistance close to the unitarity limit beyond which the model of itinerant fermions is not valid. The decrease of the resistivity of the material and/or increase in ratio of the Josephson  coupling to the charging energy, $K_0/E_c$ drives the transition with $z \to \infty$.

 Given the relationship (\ref{trans-q}) and the results in Ref. (\onlinecite{zhuchencmv}, \onlinecite{aji}), the correlation function function $\chi^H_{S^+S^-}({\bf r}, \tau)$ for the AFM, in the quantum-critical regime, is obtained from $\chi^{\tilde{H}}_{\Psi^+\Psi^-} \propto <e^{i\theta({\bf r}, \tau)} e^{-i\theta(0,0)}> $ for the XY model
\be
\label{chi}
\chi^H_{S^+S^-}({\bf r}, \tau) &=& \chi_0 \frac{1}{\tau} e^{-\sqrt{\tau/\xi_{\tau}}} \ln \big(\frac{r_c}{r}\big) e^{-r/\xi_r} e^{i{\bf Q}.{\bf r}}, \\ 
\xi_{\tau} &=& \tau_c~ e^{\sqrt{{\frac{p_c}{p_c-p}}}}; ~~ \xi_r/r_c \approx \ln (\xi_{\tau}/\tau_c).
\ee
Here $\tau$ is the imaginary time, periodic in $1/(2\pi k_BT)$, which has a lower cut-off $i\tau_c \approx (K_0/E_c)^{-1/2}$.  $p$ is the set of parameters, for example $\alpha$ and $K_0/E_c$, which drive the transition and determine the critical line $p_c$. 

There are several remarkable features in these results. The correlation function is separable in space and time; the spatial correlation length diverges only logarithmically with the temporal correlation i.e. the effective dynamical exponent $z \to \infty$; the temporal correlation at the critical point $p \to p_c$ is $1/\tau$, which gives an absorptive part as a function of $\omega$ and $T$ $\propto$ tanh$(\omega/2T)$, with an upper cut-off of order $\omega_c = (-i\tau_c)^{-1}$. This simple scaling persists over an exponentially large range in the $\big(T, (p-p_c)\big)$ plane.  

To compare with experiments, it is more useful to Fourier transform the correlation function to momentum and frequency variables. The Fourier transform to frequency space can be reduced to doing an integral which can only be evaluated numerically. The results and the fits to it to a functional form are given in Ref. (\onlinecite{zhuchencmv}). We quote this result: 
\be
\label{chi-tr}
Im ~\chi(\omega, {\bf q}) &=& - \chi_0 \tanh\Big(\frac{\omega}{2k_BT}\Big) {\mathcal{F}}_{\ell}(T \xi_{\tau}) F_c\big(\frac{\omega}{\omega_c}\big) \frac{1}{\pi} \frac{1}{|{\bf Q - q}|^2 + \kappa_{k}^{2}}, \\ \nonumber
{\mathcal{F}}_{\ell}\left(\frac{ T}{\kappa_{\omega}} \right) & \approx & \frac{1}{\left(1 +  \sqrt{ \kappa_{\omega}/2\pi T}\right)^2}, ~\text{for}~ \omega/ T \ll 1; \\ \nonumber
& \approx & \frac 14 \left(1+ 3e^{-\sqrt{\kappa_{\omega/ T}}}\right) ~\text{for}~\omega_c/T \gg \omega/T \gg 1.
\ee
$\kappa_k = \xi_r^{-1}$, and $\kappa_{\omega} = \xi_{\tau}^{-1}$ is the low frequency cut-off which increases extremely slowly (see Eq. (\ref{chi}) from 0 on deviation from criticality. $F_c\big(\frac{\omega}{\omega_c}\big)$ is a cut-off function, $F_c(0) =1, Lim(\omega >> \omega_c)~F_c\big(\frac{\omega}{\omega_c}\big) = 0.$ 
Note that $Im \chi(\omega, {\bf q})$ is a separable function of $\omega$ and $q$.
 

Since, following Caldeira-Leggett, Eqs. (\ref{af-diss}) are derived by eliminating the coupling of the collective currents to fermion currents, it follows that $\alpha = Im<j j>_{F} (q=0,\omega) = |\omega|\sigma (\omega)$. $<j j>_{F} (q=0,\omega)$ is the fermion current-current correlation in the long wave-length limit, so that $\sigma(\omega)$ is their conductivity. To test the consistency of the theory, we need to look at only the limit $\omega \to 0$, of $\sigma(0) = \rho^{-1}$, where $\rho$ is the resistivity. So, it is enough to look for the renormalization of the impurity contribution $\rho(\omega, T)$ to the resistivity. For impurities coupling to a conserved quantity, for example the density, there is no (singular) renormalization of the impurity resistivity \cite{cmv-hf}
 
{\bf Experimental Consequences}:
The results obtained in this paper are for a very simple model of itinerant Anti-ferromagnetism. The final results for the correlation function are also valid for incommensurate 2d Ising anti-ferromagnets because as discussed, their critical properties are also determined by an XY model. In heavy fermions, as well as in the Fe-based compounds, the multi-band nature of the problem and the diverse nature of the renormalization for the different orbitals with different interactions is essential for a complete description. One may ask however if universal features may govern the phenomena so that the present treatment gives some essential results. The most direct test of the applicability of the theory is a measurement of $\chi(\omega, q)$. Most critical properties can be derived once this is known.

There is only one measurement of the fluctuation spectrum at several $(q,\omega, T)$ near an AFM quantum-critical point in a quasi-2D heavy-fermion system - CeCu$_{6-x}$Au$_x$ \cite{schroder}. Within the limited accuracy of the data, taken by the essential but difficult technique of inelastic neutron scattering, the results are consistent with Eq. (\ref{chi-tr}) \cite{SVZ}, both for the $\omega/T$-dependence as well as the separability of the $\omega$ and $q$ dependence.  In the same paper \cite{SVZ}, a few results obtained \cite{Inosov} for the compound BaFe$_{1.85}$Co$_{0.15}$As$_2$ are also shown to be consistent with the results here. In neither case are the measurements done at various dopings near the critical point to study the variations with the correlation lengths. We urge more detailed experimental study of the correlation functions. It is amusing to note that the measurements on the very under-doped cuprate compounds, in the region where the AFM correlation lengths are more than about 10 lattice constants, show a frequency and temperature independent correlation length about the AFM Bragg-vectors, and a $\tanh(\omega/2T)$ scaling in $Im~\chi({\bf q}, \omega)$ \cite{AFM-cuprates}.

Earlier \cite{mfl}, one relied on the assumed non-singular nature of the spatial correlations and a momentum independent coupling vertex $g_0$ to fermions, to predict that the single-particle self-energy of the fermions, due to the interaction term $H_{int}$ is  
\be
\label{self-e}
\Sigma({\bf k}, \omega) = g_0^2 \chi_0 N(0) \Big(\omega \ln(\frac{\omega_c}{x}\big) - i\frac{\pi}{2} x\Big),
\ee
for $x \approx max(|\omega|, T) \lesssim \omega_c$. $N(0)$ is the density of states near the Fermi-energy. For $x \gtrsim \omega_c$, the imaginary part goes to a constant. The Monte-Carlo calculations have now found that the spatial correlation length also diverges, albeit only as a logarithm of the temporal correlation length, as given by Eq. (\ref{chi-tr}).  We now also have a theory of the vertex $g({\bf k,k}')$ \cite{ASV}, with which the fluctuations at momentum $({\bf k-k}')$ scatter fermions from ${\bf k}$ to ${\bf k}'$. Including both these changes, the result for the self-energy do not change in any essential way from that given by (\ref{self-e}), See Supplemental material \cite{suppl}. Given the momentum-independent self-energy, there is no back-ward scattering vertex correction for current transport. This was used in (\onlinecite{Abrahams-V-Hall}) to derive the resistivity proportional to $T$ in a solution of the Boltzmann equation including the full collision operator. The same result was obtained \cite{Shekhter-V} more formally by deriving the density-density correlation for a marginal Fermi-liquid of the conserving form with a diffusion constant proportional to $Im\Sigma $. Using the relation between the density-density and the current-current correlations, the result for the resistivity $\propto T$ is again obtained. Given such a self-energy, one can turn to the exact expression for the entropy in terms of the single-particle Green's function to find that using (\ref{self-e}), the specific heat has a singular contribution $\propto T \ln T$, except for very small T.

Both the marginal fermi-liquid energy/temperature dependence and the momentum-independence in Eq. (\ref{self-e}) are important un-tested predictions in antiferromagnetic quantum critical points. 
In multi-band compounds, such as the Fe-based high temperature superconductors, the coefficient of proportionality $g^2N(0)$ may vary between bands and be ambiguous in regions where the bands come close together. So, it is best to measure the self-energy at different angles across the various fermi-surfaces for low energies. These results are quite unlike the renormalized spin-wave theories, which has anomalous self-energies only at the "hot-points", i.e. those where the fermi-surface spans ${\bf Q}$. The results for the self-energy are much stronger than the linearity in the temperature dependence of the resistivity, which follows from it. As mentioned above, the linear in $T$ resistivity and a $T \ln T$ contribution to entropy in the quantum fluctuation regime of quasi-2D antiferromagnets appear to be universally observed. Beside the linearity in T of the resistivity, the change in resistivity in a magnetic field of the form $f(|B|/T)$, as observed \cite{Analytis}, is given by the theory because the Hamiltonian (Energy) changes linearly with $|H|$ 
through the Zeeman term and there is no linear coupling of field to the order parameter. 
It also follows \cite{mfl} from Eq. (\ref{chi-tr})  that the nuclear relaxation rate (for nuclei at which the projection of the fluctuation spectra is finite) should have a nearly constant contribution as a function of temperature, unlike the Korringa law $T_1^{-1} \propto T$ in Fermi-liquids. Evidence for such a behavior has been also found \cite{zheng-nmr} in the Fe-compounds near quantum criticality.

{\it Acknowledgements}: Very useful discussions with Elihu Abrahams, Vivek Aji, Sung-Sik Lee, Hilbert L\"ohneysen, J\"org Schmalian, Almut Schr\"oder, Frank Steglich, Qimiao Si, Peter W\"olfle and Lijun Zhu are gratefully acknowledged. This work was partially supported by the National Science Foundation grant DMR 1206298. Part of this work was done at the Aspen Center of Physics and during a Miller Professorship at University of California, Berkeley.

\newpage
{\bf Supplement: Calculation of Marginal Fermi-Liquid Self-Energy}\\

\begin{figure}[ht]
\includegraphics[width=0.5\columnwidth]{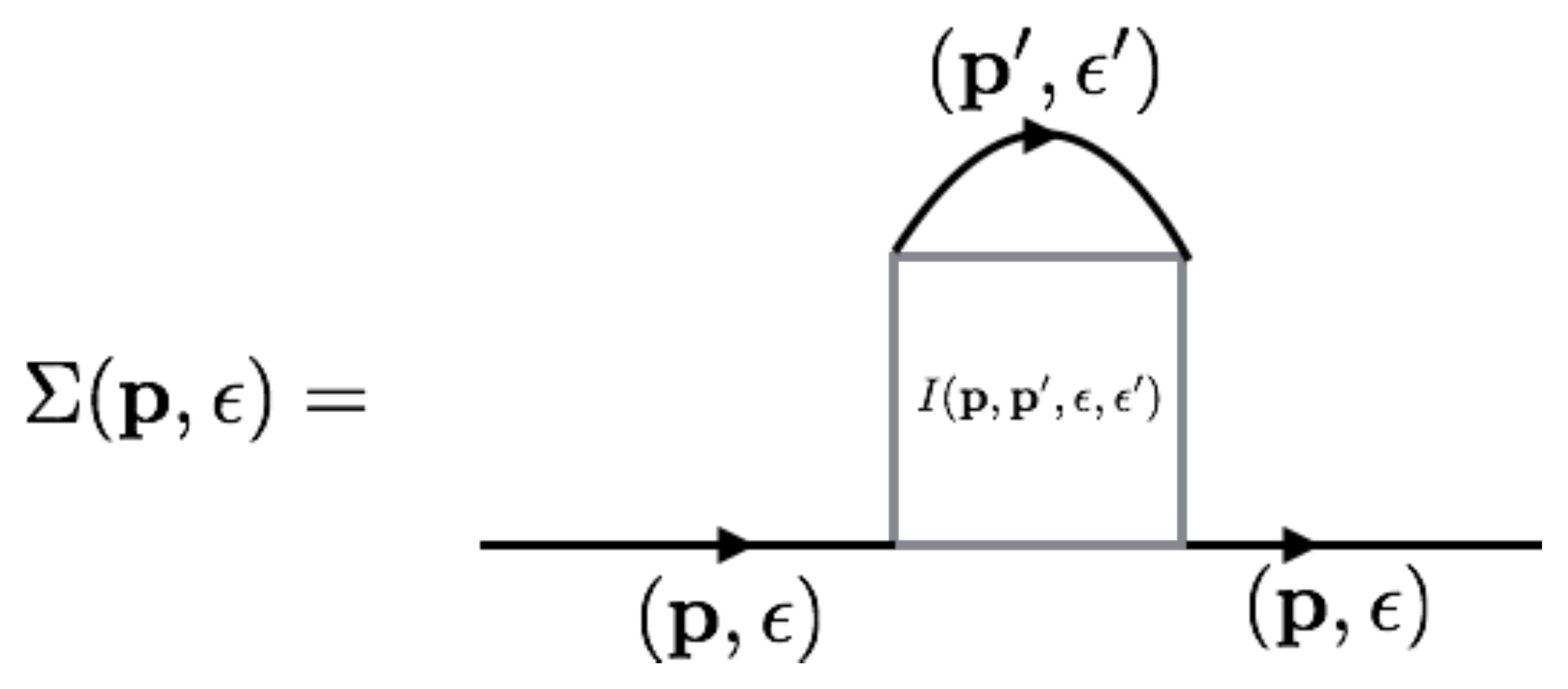}
\caption{The exact relationship of the self-energy to the irreducible vertex and the single-particle Green's function}
\label{Fig-selfen-vert}
\end{figure}

It is convenient to start with the exact relation \cite{Nozieres-book} of the one-particle self-energy $\Sigma({\bf p}, \epsilon)$ to the irreducible vertex $I({\bf p,p',q}; \epsilon, \epsilon', \nu)$ and the exact single-particle Greens' function $G({\bf p'}, \epsilon')$, as shown in Fig. (\ref{Fig-selfen-vert}). The vertex is irreducible in the particle-hole channel with total momentum-energy (${\bf q}, \nu)$ and it is assumed, as usual, that it is regular in the limit $({\bf q}, \nu) \to 0$ in this channel, which alone is needed in the self-energy calcuations. Fig. (\ref{Fig-selfen-vert}) and the associated integral equation for the self-energy given below includes all  "vertex corrections"  and self-energy insertions of the perturbative calculations. 

We are interested only in the singular contributions to the self-energy due to interactions with the collective fluctuations, specified  by Eq. (17) of the paper.  In this case, the irreducible vertex in Fig. (\ref{Fig-selfen-vert}) is proportional to the fluctuation propagator $\chi ({\bf p,p'}, \epsilon-\epsilon')$:
\be
\label{vertex}
I({\bf p,p',0}, \epsilon, \epsilon', 0) = |g({\bf p},{\bf p'})|^2 \chi({\bf p,p'}, \epsilon-\epsilon').
\ee

Following the procedure described in Ref.(\onlinecite{AGD})-sec-23.1, the self-energy shown in Fig. (\ref{Fig-selfen-vert}), with Eq. (\ref{vertex}) is given by
\be
\label{s-1}
\Sigma({\bf p}, \epsilon) &=& \frac{1}{\pi (2\pi)^d} \int d{\bf p'} |g({\bf p}, {\bf p}'|^2 \int_{-\infty}^{\infty} d\omega' \int_{-\infty}^{\infty} d\epsilon_1 \\ \nonumber
& \times & \frac{Im G_R({\bf p'}, \epsilon_1) Im \chi _R({\bf p-p'}, \omega')}{\omega' + \epsilon_1 -\omega - i\delta} \big(\tanh \frac{\epsilon_1}{2T}+ \coth \frac{\omega'}{2T}\big)
\ee
$\chi_R$ is the retarded fluctuation propagator and $G_R$ is the retarded one-particle propagator. We can follow the steps given in Ref. (\onlinecite{AGD})-sec-23.1 for evaluating the integrals in (\ref{s-1}), except that we do not assume that the imaginary part of the self-energy is insignificant as for phonons, or assume the Migdal approximation. But as in Ref. (\onlinecite{AGD}), we  assume that given the form of $\chi$, we expect the self-energy to be momentum independent. This is expected, of-course if $\chi$ were to be momentum independent, but as we will see, it is true also if $\chi$ is separable in momentum and frequency, as in Eq. (17) in the paper, of the form. Then $G({\bf p}, \epsilon)$ is given in terms of the non-interacting band-energy $\xi_{\bf p}$ and the self-energy which is to be solved for by
\be
G({\bf p}, \epsilon) = \frac{1}{\epsilon - \xi_{\bf p} - \Sigma(\epsilon)}.
\ee
Using this, we get from Eq. (\ref{s-1}) that the imaginary part of the self-energy is
\be
\label{s-2}
Im \Sigma_R({\bf p}, \epsilon) &= &\frac{\pi g_0^2}{(2\pi)^2}\frac{m}{p_F}\int_0^{k_c} dk k^2 {\cal F}_{k, }(k) \int_{-\infty}^{\infty} d\omega {\cal F}_{\omega}(\omega/2T) \big(\tanh \frac{\epsilon+\omega}{2T}+ \coth \frac{\omega}{2T}\big) \\ \nonumber
& \times &\big({\cal{T}}^{-1}(\epsilon + \omega, \xi_{|{\bf p}|+ k}) - {\cal{T}}^{-1} (\epsilon + \omega, \xi_{|{\bf p}| - k})\big).
\ee
The integrations have used  the separable form of the fluctuation propagator given by Eq. (17) in the paper and represented above by $$Im \chi _R({\bf k}, \omega') = {\cal F}_{k}({\bf k}, \kappa_r){\cal F}_{\omega}(\omega, \kappa_{\omega}).$$ Also, $|g({\bf p, p}')|^2 = g_0^2|{\bf p-p}'|^2$ derived \cite{ASV} for calculating normal state self-energy has been used.
$k_c$ is an upper-cutoff for the magnitude of momentum transfer, which is the zone-boundary, and  
\be
\label{T}
{\cal{T}}^{-1}(x,y) = \arctan\big( \frac{x- Re\Sigma(x) - y}{Im \Sigma(x)}\big); ~~ \xi_{|{\bf p}| \pm k} = \big((|{\bf p}| \pm k)^2 -p_F^2\big)/2m.
\ee
We have also specialized to 2d (although that is not necessary) and dropped a factor in the Jacobian for converting from momentum to energy integrals, which becomes important only in the region of forward scattering which is unimportant in the integral.
We expect the self-energies to be in the same scale as $\epsilon$ for $\epsilon \gtrsim T$ and on the scale of $T$ for $\epsilon \lesssim T$, i.e. smaller than the upper range $\xi(k_c)$ of the $\xi$'s. (The calculation below does not change if there are logarithmic correction to $Re \Sigma (\epsilon)$). Given the range of the $k$-integral, the restrictions on the 
$\omega$-integral from the ${\cal{T}}$ factors is over the band-width $\xi(k_c)\pm \Sigma(\epsilon)$ corrections. The corrections due to 
$\Sigma(\epsilon)$ are un-important for $\epsilon$ of interest because the range of $\omega$ integration is actually
 limited by the thermal factors in (\ref{s-2}) to the much smaller energies of $O\big(max(\epsilon, T)\big)$. 
The upper limit on the integral over $k$ can therefore be done easily over its entire range. We are left only with the $\omega$ integral. Now we note that in the quantum-critical regime, the temporal corelation length in Eq. (17) of the paper $\xi_{\tau} << T$, so that $F(\omega) = -\chi_0 \tanh{(\omega/2T)}$. In this regime the self-energy is then given by
\be
Im \Sigma_R({\bf p}, \epsilon) &=& \overline{g}_0^2 N(0) \chi_0 max(|\epsilon|, T), ~\text{for}~ max(||\epsilon|, T) \lesssim \omega_c, \\ \nonumber
&= & \overline{g_0}^2 N(0) \chi_0 \omega_c, ~\text{for} ~max(|\epsilon|, T) \gtrsim \omega_c
\ee
$\overline{g_0}$ includes numerical corrections of O(1) to $g_0$, which depend on details of the band-structure. 

For the regime, $\kappa_{\omega} >> T$, the integral over $\omega$ is cut-off by $\kappa_{\omega}$ and the contribution to self-energy becomes $\omega^2 /\kappa_{\omega}$  which  vanishes as one deviates far from the critical point. This adds to the normal non-singular Fermi-liquid scattering which is always present.

These results are identical in functional form to the perturbative results. That they are true more generally was stated without proof in Ref. (\onlinecite{Kotliar-epl}) and the relations of the irreducible vertex to the complete vertex and to density-density correlations in the hydrodynamic regime were derived in Ref. (\onlinecite{Shekhter-V}).What is new in this note is that the same form of the results is true for separable collective fluctuations as local ($q$-independent) fluctuations.

\end{document}